\documentstyle[12pt]{article}
\begin{document}

\begin{flushright}{UT-03-03}
\end{flushright}
\vskip 0.5 truecm

\begin{center}
{\Large{\bf Lattice chiral symmetry, CP-violation and Majorana 
fermions }}
\end{center}
\vskip .5 truecm
\centerline{\bf Kazuo Fujikawa}
\vskip .4 truecm
\centerline {\it Department of Physics,University of Tokyo}
\centerline {\it Bunkyo-ku,Tokyo 113,Japan}
\vskip 0.5 truecm

%\makeatletter
%\@addtoreset{equation}{section}
%\def\theequation{\thesection.\arabic{equation}}
%\makeatother

\begin{abstract}
A brief summary of lattice fermions defined by  
the general Ginsparg-Wilson algebra is first given.
It is then shown that those general class of fermion operators
have a conflict with CP invariance in chiral gauge theory and 
with the definition of Majorana
fermions in the presence of chiral-symmetric Yukawa couplings.
The same conclusion holds for the domain-wall fermion also. 
\end{abstract}

\section{Introduction}

The recent developments in the analysis of chiral symmetry in
lattice theory is based on the so-called  Ginsparg-Wilson 
relation\cite{ginsparg}. Neuberger found an explicit 
construction of lattice 
fermion operator (the overlap operator)\cite{neuberger}, which 
was later shown to
satisfy the Ginsparg-Wilson relation. This operator was then 
shown to satisfy an index theorem on the lattice under certain 
conditions\cite{hasenfratz3}.  This index relation was used by 
L\"uscher to 
derive the chiral anomaly as a Jacobian factor\cite{luscher} 
just as in continuum theory\cite{fujikawa}. 
These topological properties were later examined in
further detail\cite{ky}. See 
Refs.\cite{niedermayer} for reviews of these developments.

To be more specific, we here work on the Ginsparg-Wilson 
relation defined by\cite{fujikawa2}   
\begin{eqnarray}
\gamma_{5}(\gamma_{5}D)+(\gamma_{5}D)\gamma_{5}=
2a^{2k+1}(\gamma_{5}D)^{2k+2}
\end{eqnarray}
where $D$ is the lattice Dirac operator and $k$ stands for a 
non-negative integer; $k=0$ corresponds to the ordinary 
Ginsparg-Wilson relation. 

We deal with a hermitian lattice operator 
\begin{eqnarray}
H=a\gamma_{5}D=H^{\dagger}=aD^{\dagger}\gamma_{5}
\end{eqnarray}
and the above algebra is written as  
\begin{eqnarray}
\gamma_{5}H+H\gamma_{5}=2H^{2k+2}
\end{eqnarray}
where  $k=0$ corresponds to
conventional Ginsparg-Wilson relation.
We also assume that the operator $H$ is local in the sense that 
it is analytic in the entire Brillouin zone and free of species 
doublers, which is in fact 
shown for the free operator without gauge field\cite{fi2}. 
One can then confirm the relation
\begin{eqnarray}
\gamma_{5}H^{2}=H^{2}\gamma_{5}.
\end{eqnarray} 
The defining algebra is written in  various ways such as  
\begin{eqnarray}
&&\Gamma_{5}H+H\Gamma_{5}=0,\nonumber\\
&&\gamma_{5}H+H\hat{\gamma}_{5}=0,\nonumber\\
&&\hat{\gamma}^{2}_{5}=1
\end{eqnarray}
where
\begin{eqnarray}
&&\Gamma_{5}=\gamma_{5}-H^{2k+1},\nonumber\\
&&\hat{\gamma}_{5}=\gamma_{5}-2H^{2k+1}.
\end{eqnarray}
We can also show
\begin{eqnarray}
(\gamma_{5}\Gamma_{5})\hat{\gamma}_{5}
&=&\gamma_{5}(\gamma_{5}\Gamma_{5}).
\end{eqnarray}

We now examine the action defined by
\begin{eqnarray}
S=\int d^{4}x\bar{\psi}D\psi\equiv\sum_{x,y}\bar{\psi}(x)D(x,y)
\psi(y)
\end{eqnarray}
which is invariant under
\begin{eqnarray}
\delta\psi=i\epsilon\hat{\gamma}_{5}\psi, \ \ 
\delta\bar{\psi}=\bar{\psi}i\epsilon\gamma_{5}.
\end{eqnarray}
If one considers the field re-definition
\begin{eqnarray}
\psi^{\prime}=\gamma_{5}\Gamma_{5}\psi,\ \ \ 
\bar{\psi}^{\prime}=\bar{\psi}
\end{eqnarray}
the above action is written as 
\begin{eqnarray}
S=\int d^{4}x\bar{\psi}^{\prime}D\frac{1}{\gamma_{5}\Gamma_{5}}
\psi^{\prime}
\end{eqnarray}
which is invariant under naive chiral transformation 
\begin{eqnarray}
\delta\psi^{\prime}&=&
i\epsilon\gamma_{5}\psi^{\prime},
\nonumber\\
\delta\bar{\psi}^{\prime}&=&
\bar{\psi}^{\prime}i\epsilon
\gamma_{5}.
\end{eqnarray}
This chiral symmetry implies the relation
\begin{eqnarray}
\{\gamma_{5},D\frac{1}{\gamma_{5}\Gamma_{5}} \}=0.
\end{eqnarray}
This naive chiral symmetry of the species doubler-free operator
suggests the non-analytic behavior of the factor  
$1/(\gamma_{5}\Gamma_{5})$ in 
the Brillouin zone. In fact, one can confirm that\cite{fi2,fi1}  
\begin{eqnarray}
\Gamma^{2}\equiv \Gamma^{2}_{5}=1-H^{4k+2}=0
\end{eqnarray}
has solutions at the momentum variables corresponding to 
would-be species doublers, in the case of free operator without
the gauge field, and also in 
the presence of topologically non-trivial gauge field.
See also Ref.\cite{chiu}.

We next recall the charge conjugation properties
\begin{eqnarray}
&&C\gamma^{\mu}C^{-1}=-(\gamma^{\mu})^{T},\ \  
C\gamma_{5}C^{-1}=\gamma^{T}_{5},\nonumber\\ 
&&C^{\dagger}C=1,\ \  C^{T}=-C.
\end{eqnarray}
We then have\footnote{We define the CP operation by 
$W=C\gamma_{0}=\gamma_{2}$ with hermitian $\gamma_{2}$ and 
the CP 
transformed gauge field by $U^{\rm CP}$, and then 
$WD(U^{\rm CP})W^{-1}=D(U)^T$. If the parity is realized in the 
standard way, we have $CD(U^{\rm C})C^{-1}=D(U)^T$.} 
\begin{eqnarray}
&&WD(U^{\rm CP})W^{-1}=D(U)^T,\qquad
W\gamma_{5}\Gamma_{5}(U^{\rm CP})W^{-1}=[\gamma_{5}
\Gamma_{5}(U)]^T,
\nonumber\\
&&WH(U^{\rm CP})W^{-1}=-[\gamma_{5}H(U)\gamma_{5}]^T,\qquad
WH^{2}(U^{\rm CP})W^{-1}=[H^{2}(U)]^T,
\nonumber\\
&&W\Gamma_{5}(U^{\rm CP})W^{-1}=-[\gamma_{5}\Gamma_{5}(U)
\gamma_{5}]^T,
\nonumber\\
&&W(\Gamma_{5}/\Gamma)(U^{\rm CP})W^{-1}
=-[(\gamma_{5}\Gamma_{5}\gamma_{5}/\Gamma)(U)]^T
\end{eqnarray}
where 
\begin{equation}
\Gamma=\sqrt{\Gamma^{2}_{5}}
=\sqrt{(\gamma_{5}\Gamma_{5}\gamma_{5})^{2}}
=\sqrt{1-H^{4k+2}}.
\end{equation}
Here we imposed the relation $WD(U^{\rm CP})W^{-1}=D(U)^T$ 
or~$[CD(U)]^T=-CD(U^{\rm C})$ which is consistent with the 
defining Ginsparg-Wilson relation.

We also have the property
\begin{eqnarray}
&&W\hat{\gamma}_{5}(U^{\rm CP})W^{-1}
=-\left[\gamma_{5}\hat{\gamma}_{5}(U)\gamma_{5}\right]^{T}.
\end{eqnarray}

\section{CP symmetry in lattice chiral gauge theory}

We now examine the CP symmetry in chiral gauge theory 
\begin{equation}
{\cal L}_{L}=\bar{\psi}_{L}D\psi_{L}
\end{equation}
where we defined the (general) projection operators 
\begin{eqnarray}
&&D=\bar{P}_{L}DP_{L}+\bar{P}_{R}DP_{R},
\nonumber\\
&&\psi_{L,R}=P_{L,R}\psi,\qquad\bar{\psi}_{L,R}
=\bar{\psi}\bar{P}_{L,R}. 
\end{eqnarray} 
It was pointed out by Hasenfratz\cite{hasenfratz} that 
the conventional Ginsparg-Wilson operator when applied to 
chiral gauge theory has a difficulty with CP symmetry. We would 
like to examine this issue in more detail.
Under the standard CP transformation\footnote{The vector-like 
theory is invariant under this CP transformation.} 
\begin{eqnarray}
&&\bar{\psi}\rightarrow \psi^{T}W,
\nonumber\\
&&\psi\rightarrow -W^{-1}\bar{\psi}^{T}
\end{eqnarray}
the chiral action is invariant only if
\begin{eqnarray}
&&W P_{L}W^{-1}=\bar{P}^{T}_{L},\qquad W\bar{P}_{L}W^{-1}
=P^{T}_{L}. 
\end{eqnarray} 
It was shown in Ref.~\cite{fis1} that the unique solution for 
this condition in the framework of the Ginsparg-Wilson 
operators is given by 
\begin{eqnarray}
P_{L,R}=\frac{1}{2}(1\mp\Gamma_5/\Gamma),
\nonumber\\
\bar{P}_{L,R}=\frac{1}{2}(1\pm\gamma_5\Gamma_5\gamma_5/\Gamma),
\end{eqnarray}
but these projection operators suffer from singularities 
in~$1/\Gamma$, as we have already noted. Namely, it is 
impossible to maintain the manifest CP invariance of the local 
and chiral symmetric doubler-free 
Lagrangian~\cite{hasenfratz,fis1,fis2}.

If one stays in the well-defined local Lagrangian 
\begin{equation}
\int{\cal L}_{L}=\int\bar{\psi}P_{+}D\hat{P}_{-}\psi
\end{equation}
where
\begin{eqnarray}
&&P_{\pm}=\frac{1}{2}(1\pm\gamma_{5}),\nonumber\\
&&\hat{P}_{\pm}=\frac{1}{2}(1\pm\hat{\gamma}_{5})
\end{eqnarray}
it is not invariant under the standard CP transformation as 
\begin{eqnarray}
&&WP_{\pm}W^{-1}=P^{T}_{\mp}\neq\hat{P}^{T}_{\mp}(U),
\nonumber\\
&&W\hat{P}_{\pm}(U^{\rm CP})W^{-1}
=\frac{1\mp\left[\gamma_{5}\hat{\gamma}_{5}(U)
\gamma_{5}\right]^T}{2}
=\left[\gamma_{5}\hat{P}_{\mp}(U)\gamma_{5}\right]^{T}
\neq P^{T}_{\mp}, 
\nonumber\\
&&\left[WP_{+}D(U^{\rm CP})\hat{P}_{-}(U^{\rm CP})W^{-1}
\right]^{T}
=\gamma_{5}\hat{P}_{+}(U)\gamma_{5}D(U)P_{-}
\nonumber\\
&&=P_{+}D(U)\hat{P}_{-}(U)-D(U)[\gamma_{5}-\Gamma_{5}(U)]
\neq P_{+}D\hat{P}_{-}. 
\end{eqnarray} 
This generalizes the analysis of Hasenfratz in 
a more general setting\footnote{This analysis is extended to a 
more general class of operators defined by 
$\gamma_{5}H+H\gamma_{5}=2H^{2}f(H^{2})$, where $f(H^{2})$
is a regular and monotonous non-decreasing function\cite{fis1}
. }.

\section{ Majorana fermion}
We next want to show that the general class of Ginsparg-Wilson 
operators have a difficulty to define Majorana fermions in the 
presence of chiral symmetric Yukawa couplings\cite{fi1}.
We start with
\begin{eqnarray}
{\cal L}
&=&\bar{\psi}_{R}D\psi_{R}+\bar{\psi}_{L}D\psi_{L} 
+ m[\bar{\psi}_{R}\psi_{L}+\bar{\psi}_{L}\psi_{R}]\nonumber\\
&&+ 2g[\bar{\psi}_{L}\phi\psi_{R}
+\bar{\psi}_{R}\phi^{\dagger}\psi_{L}]\nonumber\\
&=&\bar{\psi}D\psi 
+ m\bar{\psi}\gamma_{5}\Gamma_{5}\psi\nonumber\\
&&+ \frac{g}{\sqrt{2}}\bar{\psi}[A+
(\gamma_{5}\Gamma_{5}\gamma_{5}/\Gamma)A
(\Gamma_{5}/\Gamma)
+i(\gamma_{5}\Gamma_{5}\gamma_{5}/\Gamma)B
+iB(\Gamma_{5}/\Gamma)]\psi
\end{eqnarray}
where
\begin{equation}
\psi_{L,R}=P_{L,R}\psi,\qquad\bar\psi_{L,R}
=\bar\psi\bar P_{L,R}
\end{equation}
with the projection operators in (23),
and  we used $\phi=(A + iB)/\sqrt{2}$. 
We then  make the substitution\cite{nicolai, van nieuwenhuizen} 
\begin{eqnarray}
&&\psi=(\chi+i\eta)/\sqrt{2},\nonumber\\
&&\bar{\psi}=(\chi^{T}C-i\eta^{T}C)/\sqrt{2}
\end{eqnarray}
and obtain
\begin{eqnarray}
{\cal L}
&=&\frac{1}{2}\chi^{T}CD\chi 
+ \frac{1}{2}m\chi^{T}C\gamma_{5}\Gamma_{5}\chi\nonumber\\
&&+ \frac{g}{2\sqrt{2}}\chi^{T}C[A+
(\gamma_{5}\Gamma_{5}\gamma_{5}/\Gamma)A
(\Gamma_{5}/\Gamma)
+i(\gamma_{5}\Gamma_{5}\gamma_{5}/\Gamma)B
+iB(\Gamma_{5}/\Gamma)]\chi\nonumber\\
&+&\frac{1}{2}\eta^{T}CD\eta
+ \frac{1}{2}m\eta^{T}C\gamma_{5}\Gamma_{5}\eta\nonumber\\
&&+ \frac{g}{2\sqrt{2}}\eta^{T}C[A+
(\gamma_{5}\Gamma_{5}\gamma_{5}/\Gamma)A
(\Gamma_{5}/\Gamma)
+i(\gamma_{5}\Gamma_{5}\gamma_{5}/\Gamma)B
+iB(\Gamma_{5}/\Gamma)]\eta.
\end{eqnarray} 
This relation shows that we can write the Dirac fermion 
operator as a sum of two Majorana operators.
One can then define the Majorana fermion $\chi$ (or $\eta$) and 
the resulting Pfaffian as a square root of the determinant of
lattice Dirac operator.
But this formulation  of the Majorana fermion
inevitably suffers from the singularities of the modified 
chiral operators $\Gamma_{5}/\Gamma$ and 
$\gamma_{5}\Gamma_{5}\gamma_{5}/\Gamma$ in the Brillouin zone,
as we have already explained.

We note that the condition,
\begin{eqnarray}
&&CP_{L}C^{-1}=\bar{P}^{T}_{R}, \ \ \ C\bar{P}_{L}C^{-1}
=P^{T}_{R}
\end{eqnarray}
which is required by the consistent $CP$ property, is directly 
related to the condition of the consistent Majorana reduction 
for the term containing scalar field  $A(x)$,
\begin{eqnarray} 
&&C(\gamma_{5}\Gamma_{5}\gamma_{5}/\Gamma)A(x)
(\Gamma_{5}/\Gamma)
=-[C(\gamma_{5}\Gamma_{5}\gamma_{5}/\Gamma)
A(x)(\Gamma_{5}/\Gamma)]^{T}
\end{eqnarray}
in the Yukawa coupling\footnote{If $(CO)^{T}=-CO$ 
for a general operator $O$, the cross
term vanishes $\eta^{T}CO\chi-\chi^{T}CO\eta=0$ by using the 
anti-commuting property of $\chi$ and $\eta$.}, if one recalls that 
the difference 
operators in $\Gamma_{5}$ and $\Gamma$ do not commute with 
the field $A(x)$.\\

 In other words, if one uses the projection 
operators which do not satisfy,
\begin{eqnarray}
&&CP_{L}C^{-1}=\bar{P}^{T}_{R}, \ \ \ C\bar{P}_{L}C^{-1}
=P^{T}_{R}
\end{eqnarray}
the consistent Majorana reduction is {\em not} realized. 
For the chiral symmetric Yukawa
couplings such as in supersymmetry  the Majorana reduction is 
thus directly related 
to the condition of the CP invariance,  
provided that parity properties are the standard 
ones.

\section{Discussion and conclusion}
 
We have shown that both of 
the consistent definitions of CP symmetry in chiral gauge 
theory and the Majorana fermion in the 
presence of chiral symmetric Yukawa couplings 
are based on the same condition,
\begin{eqnarray}
&&CP_{L}C^{-1}=\bar{P}^{T}_{R}, \ \ \ C\bar{P}_{L}C^{-1}
=P^{T}_{R}
\end{eqnarray}
and that the construction of projection operators, which are
the {\em unique solution},
\begin{eqnarray}
P_{L,R}=\frac{1}{2}(1\mp \Gamma_{5}/\Gamma),\nonumber\\
\bar{P}_{L,R}
=\frac{1}{2}(1\pm \gamma_{5}\Gamma_{5}\gamma_{5}/\Gamma)
\end{eqnarray} 
inevitably suffers from singularities in the modified chiral 
operators (to be precise, in $1/\Gamma$) for any Dirac operator 
$D$ satisfying the algebraic relation 
\begin{eqnarray}
\gamma_{5}H+H\gamma_{5}=2H^{2k+2}
\end{eqnarray}
with $H=a\gamma_{5}D$.  
We find it interesting that the breaking of CP
symmetry and a conflict with Majorana reduction are directly 
related to the basic notions of locality and species doubling
in lattice theory.

If one uses well-defined regular lattice operators
\begin{eqnarray}
\int{\cal D}\psi_{L}{\cal D}\bar{\psi}_{L}
\exp\left(\int\bar{\psi}P_{+}D\hat{P}_{-}\psi\right)
\end{eqnarray}
with
\begin{eqnarray}
&&P_{\pm}=\frac{1}{2}(1\pm\gamma_{5}),\nonumber\\
&&\hat{P}_{\pm}=\frac{1}{2}(1\pm\hat{\gamma}_{5})
\end{eqnarray}
the CP breaking in lattice chiral theory on the basis of 
Ginsparg-Wilson operator inevitably appears. 
As for the physical implications of this breaking of CP
symmetry in pure lattice chiral theory, it was shown in 
Ref.\cite{fis2} following the formulation in\cite{luscher2,
suzuki,
luscher3, suzuki2}
 that the effects of CP breaking are isolated in the (almost) 
contact term of fermion propagator, which is connected to the 
external fermion sources. The CP breaking effects in all the 
loop diagrams are under well-control and they are absorbed into
the weight factors related to various topological sectors in the 
fermionic path integral.
In the presence of the Higgs coupling and in particular in the 
presence of the vacuum expectation value of the Higgs field,
the analysis of CP breaking becomes more involved and it could be
more serious\cite{fis2}.  

It is also shown that this complication of CP breaking persists 
in the domain-wall fermion\cite{kaplan, shamir, vranas, 
neuberger2, kikukawa2}. In fact, it is shown that the 
domain-wall fermion in the limit $N=\infty$, when applied to 
chiral theory, is valid only for the topologically trivial 
sector
 and that it still suffers from the complications related to 
CP symmetry\cite{fs}. In the $N=\infty$ limit, the domain-wall
fermion is related to the Ginsparg-Wilson fermion by 
\begin{eqnarray}
&&\int{\cal D}\psi_{L}{\cal D}\bar{\psi}_{L}
\exp\left(\int\bar{\psi}P_{+}D\hat{P}_{-}\psi\right)
\nonumber\\
&&=\int{\cal D}q_{L}{\cal D}\bar{q}_{L}{\cal D}Q_{L}
{\cal D}\bar{Q}_{R}
\exp\left(\int\bar{q}P_{+}D\frac{1}{\gamma_{5}\Gamma_{5}}P_{-}q
+\int\bar{Q}\hat{P}_{-}\frac{1}{\gamma_{5}\Gamma_{5}}P_{-}
Q\right)
\end{eqnarray}
where $q$ and $\bar{q}$ stand for the standard variables in 
domain-wall fermion, and $Q$ and $\bar{Q}$ stand for the 
Pauli-Villars fields. Note the appearance of the non-local
operator $D/(\gamma_{5}\Gamma_{5})$.

It is customary to use the domain-wall fermion for finite 
$N$ in practical applications, but we consider that the finite 
$N$ theory is not in a better situation with respect to CP 
symmetry either\footnote{One may argue that the domain 
wall variables $q_{L}$ and~$\bar q_{L}$, which become non-local 
and cannot 
describe topological properties in the limit~$N=\infty$, are not
 the suitable variables to describe physical correlation 
functions even for finite $N$, to the extent that the finite 
$N$ theory is intended to be an approximation to the theory with
 $N=\infty$.}, besides the ill-defined chiral symmetry.


\begin{thebibliography}{99}

\bibitem{ginsparg}
P.H. Ginsparg and K.G. Wilson,
Phys. Rev. D~25 (1982) 2649.
\bibitem{neuberger}
H. Neuberger,
Phys.\ Lett.\ B~417 (1998) 141;
Phys.\ Lett.\ B~427 (1998) 353.
\bibitem{hasenfratz3}
P. Hasenfratz, V. Laliena and F. Niedermayer,
Phys.\ Lett.\ B~427 (1998) 125.
\bibitem{luscher}
 M. L\"uscher,
 Phys.\ Lett.\ B~428 (1998) 342.
\bibitem{fujikawa}
 K. Fujikawa,
 Phys.\ Rev.\ Lett.\ 42 (1979) 1195;
 Phys.\ Rev.\ D~21 (1980) 2848;
 [Erratum-ibid.\ D~22 (1980) 1499].
\bibitem{ky}
 Y. Kikukawa and A. Yamada,
 Phys.\ Lett.\ B~448 (1999) 265;\\
 D.H. Adams, Annals Phys.\  296 (2002) 131;\\
 H. Suzuki,
 Prog.\ Theor.\ Phys.\ 102 (1999) 141;\\ 
 K. Fujikawa, Nucl.\ Phys.\ B~546 (1999) 480.
\bibitem{niedermayer}
 F. Niedermayer, Nucl.\ Phys.\ (Proc. Suppl.) 73 (1999) 105;\\ 
 H. Neuberger, Ann.\ Rev.\ Nucl.\ Part.\ Sci.\ 51 (2001) 23;\\ 
 M. L\"uscher, lectures given at the International School of 
Subnuclear physics, Erice 2000, hep-th/0102028;\\ 
Y. Kikukawa, Nucl.\ Phys.\ (Proc. Suppl.) 106 (2002) 71. 
\bibitem{fujikawa2}
K. Fujikawa,
Nucl.\ Phys.\ B~589 (2000) 487.
\bibitem{fi2}
K. Fujikawa and M. Ishibashi,
Nucl.\ Phys.\ B~587 (2000) 419;
Nucl.\ Phys.\ B~605 (2001) 365.
\bibitem{fi1}
K. Fujikawa and M. Ishibashi,
Nucl.\ Phys.\ B~622 (2002) 115;
Phys.\ Lett.\ B~528 (2002) 295.
\bibitem{chiu}
T.W. Chiu, C.W. Wang and S.V. Zenkin,
Phys.\ Lett.\ B~438 (1998) 321.
\bibitem{hasenfratz}
P. Hasenfratz,
Nucl.\ Phys.\ (Proc.\ Suppl.) 106 (2002) 159, 
and references therein. 
\bibitem{fis1}
K. Fujikawa, M. Ishibashi and H. Suzuki,
Phys.\ Lett.\ B~538 (2002) 197.
\bibitem{fis2}
K. Fujiwawa, M. Ishibashi and H. Suzuki,
J. High Energy Phys. 04 (2002) 046.
\bibitem{nicolai}
H. Nicolai,
Nucl.\ Phys.\ B~140 (1978) 294;
Nucl.\ Phys.\ B~156 (1979) 157.
\bibitem{van nieuwenhuizen}
P. van Nieuwenhuizen and A. Waldron,
Phys.\ Lett.\ B~389 (1996) 29.
\bibitem{luscher2}
M. L\"uscher,
Nucl.\ Phys.\ B~549 (1999) 295.
\bibitem{suzuki}
H. Suzuki,
Prog.\ Theor.\ Phys.\ 101 (1999) 1147. 
\bibitem{luscher3}
M. L\"uscher,
Nucl.\ Phys.\ B~568 (2000) 162.
\bibitem{suzuki2}
H. Suzuki,
Nucl.\ Phys.\ B~585 (2000) 471;\\
H. Igarashi, K. Okuyama and H. Suzuki,
hep-lat/0012018.
\bibitem{kaplan}
D.B. Kaplan,
Phys.\ Lett.\ B~288 (1992) 342.
\bibitem{shamir}
Y. Shamir,
Nucl.\ Phys.\ B~406 (1993) 90;\\
V. Furman and Y. Shamir,
Nucl.\ Phys.\ B~439 (1995) 54.
\bibitem{vranas}
P. Vranas,
Phys.\ Rev.\ D~57 (1998) 1415.
\bibitem{neuberger2}
H. Neuberger,
Phys.\ Rev.\ D~57 (1998) 5417;
Phys.\ Rev.\ D~59 (1999) 085006.
\bibitem{kikukawa2}
Y. Kikukawa and T. Noguchi,
hep-lat/9902022.
\bibitem{fs}
K. Fujikawa and H, Suzuki, hep-lat/0210013 (To appear in 
Phys. Rev. D).
\end{thebibliography}
\end{document}